\documentclass[prl,twocolumn,superscriptaddress,longbibliography]{revtex4-2}
\usepackage{amsmath}
\usepackage{amssymb}
\usepackage{amsthm}
\usepackage{amsfonts}
\usepackage{listings}
\usepackage{enumerate}
\usepackage{latexsym}
\usepackage{psfrag}
\usepackage{bm} 
\usepackage[all]{xy}
\usepackage{subfigure}
\usepackage{color}
\usepackage{mathtools}
\usepackage{array}
\usepackage{placeins}
\usepackage{hyperref}
\usepackage[export]{adjustbox}

\usepackage{physics}
\usepackage{tikz}
\usetikzlibrary{calc}
\usetikzlibrary{decorations.pathreplacing}

\begin{document}

\title{Simulating Bulk Gap in Chiral Projected Entangled-Pair States}

\author{Ji-Yao Chen}
\thanks{These two authors contribute equally.}
\affiliation{Center for Neutron Science and Technology, Guangdong Provincial Key Laboratory of Magnetoelectric Physics and Devices, School of Physics, Sun Yat-sen University, Guangzhou 510275, China}

\author{Yi Tan}
\thanks{These two authors contribute equally.}
\affiliation{Department of Physics, Southern University of Science and Technology, Shenzhen 518055, China}

\author{Sylvain Capponi}
\author{Didier Poilblanc}
\affiliation{Laboratoire de Physique Th\'eorique, Universit\'e de Toulouse, CNRS, UPS, France}

\author{Fei Ye}
\author{Jia-Wei Mei}
\email{meijw@sustech.edu.cn}
\affiliation{Department of Physics, Southern University of Science and Technology, Shenzhen 518055, China}

\date{\today}

\begin{abstract}
Projected entangled-pair states (PEPS) have proven effective in capturing chiral spin liquid ground states, yet the presence of long-range ``gossamer'' correlation tails raises concerns about their ability to accurately describe bulk gaps. Here, we address this challenge and demonstrate that PEPS can reliably characterize gapped bulk excitations in chiral topological phases. Using a variational principle for excited states within a local mode approximation, we establish that correlation functions decaying faster than $r^{-2}$ are not necessarily related to gapless modes and thus long-range ``gossamer'' correlation tails in chiral PEPS do not contradict the presence of a bulk gap. This framework is validated in the spin-$\frac{1}{2}$ Kitaev model with a chiral term, where PEPS yields excitation gaps that agree well with exact solutions. Extending our approach to the $\mathbb{Z}_3$ Kitaev model, we present compelling evidence for its chiral ground state and accurately resolve its gapped excitations. These findings thus solidify PEPS as a powerful tool for studying both ground and excited states in chiral topological systems, thereby bridging a key gap in the understanding of their bulk properties.
\end{abstract}
\maketitle

\emph{Introduction. --}
Chiral spin liquids (CSLs) are two-dimensional lattice analogs of fractional quantum Hall states~\cite{Laughlin1983, Kalmeyer1987, Wen1989a} in spin systems, characterized by topological order~\cite{Wen1990a}. While projected entangled-pair states (PEPS), a powerful tensor-network method, have demonstrated remarkable success in capturing nonchiral topological phases~\cite{Verstraete2004,Schuch2010,Cirac2021}, their extension to chiral phases poses significant challenges.
Align with the no-go theorem for chiral free-fermionic states~\cite{Dubail2015}, initial studies demonstrated that Gaussian fermionic PEPS could efficiently approximate the topological features of Chern insulators, though their correlation functions exhibit algebraic decay~\cite{Wahl2013,Wahl2014,Dubail2015}. Subsequently, SU(2)-invariant PEPS and their SU(N) generalizations~\cite{Poilblanc2016,Poilblanc2017b,Chen2018,Chen2020}, designed to simulate CSL Hamiltonians, were found to exhibit slower-than-exponential, ``long-range'' correlation decay. Similar behaviors have also been observed in unrestricted PEPS studies of chiral topological phases~\cite{Hasik2022,Weerda2024} and in Gutzwiller-projected Gaussian PEPS~\cite{Yang2015a,Niu2024}.  
The persistent long-range ``gossamer'' tails—characterized by correlations of very small magnitude—raise critical concerns about whether PEPS can accurately capture the bulk properties of CSLs, particularly their gapped bulk excitations.

More specifically, this skepticism stems from the well-established correspondence between the correlation length and the energy gap, a fundamental diagnosis in many-body systems~\cite{Anderson1997,Chaikin1995}. In gapped phases, correlation functions exhibit exponential decay with distance, reflecting the finite energy cost of excitations and the localized nature of disturbances within the system. By contrast, gapless systems are characterized by algebraic decay, indicative of long-range correlations associated with low-energy, long-wavelength excitations whose energies approach that of the ground state. Generally, the relationship between the correlation length \(\xi\) and the energy gap \(\Delta\) follows the inverse scaling \(\Delta \sim 1/\xi\), such that the diverging \(\xi\) signals a vanishing \(\Delta\). Applying this framework to chiral PEPS, the presence of long-range ``gossamer'' tails suggests the absence of a well-defined bulk gap, raising significant concerns about its reliability as a tool for describing chiral topological order.

In this work, we demonstrate that PEPS can faithfully simulate gapped bulk excitations of CSL, despite their diverging correlation length.
Based on recently developed tangent space based excitation ansatz~\cite{Vanderstraeten2015,Vanderstraeten2019a,Ponsioen2020,Ponsioen2022,Ponsioen2023b,Tu2023a}, a tensor network generalization of
local mode approximation~\cite{Feynman1953a}, we derive a variational principle for excited-state energies and establish that correlation functions decaying faster than $r^{-2}$ do not inherently lead to unphysical gapless bulk states.
We implement 1-form symmetric PEPS~\cite{Tan2024, Tan2024a} to study \(\mathbb{Z}_N^{[1]}\) (\(N = 2, 3\)) 1-form symmetric Kitaev-type models~\cite{Kitaev2006a, Barkeshli2015, Chen2024, Ellison2023,Ma2023, Liu2024}, which host chiral ground states in the phase diagram~\cite{Kitaev2006a, Chen2024}. 
Among these models, the spin-\(\frac{1}{2}\) Kitaev model with a chiral term offers an exact solution~\cite{Kitaev2006a,Lahtinen2008}, providing a rigorous benchmark to validate the precision of our PEPS simulations in capturing both ground and excited state properties. Then we extend our framework to the \(\mathbb{Z}_3\) Kitaev model with \(\mathbb{Z}_3^{[1]}\) 1-form symmetry~\cite{Barkeshli2015, Ellison2023, Chen2024}, a system that lacks an exact solution. In this case, we successfully confirm the chiral nature of its ground state and accurately compute its gapped bulk excitations, further reinforcing the robustness and versatility of our approach.

\emph{Variational principle for excited states. --}
In this work, we focus on the honeycomb lattice, though our method generalizes naturally to other two-dimensional lattices. The translationally invariant PEPS for the honeycomb lattice is expressed as 
\begin{eqnarray} 
|\Psi(A)\rangle = \sum_{s_{\alpha_i},\cdots,s_{\beta_N}}{\rm tTr} (A_{abc}^{s_{\alpha_i}} A_{a'b'c}^{s_{\beta_i}}) |s_{\alpha_1} \cdots s_{\alpha_i}s_{\beta_i}\cdots s_{\beta_N}\rangle\nonumber,
\end{eqnarray}
represented by the graph $\includegraphics[width=0.45\columnwidth,valign=c]{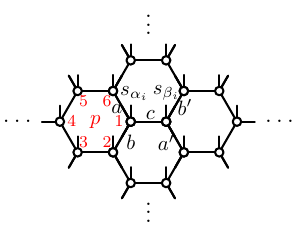}$. The local tensor \(A_{abc}^{s}\) includes one physical index (\(s\)) and three virtual indices (\(a\), \(b\), and \(c\)) corresponding to the nearest-neighbor bonds.

While the set of uniform PEPS defines a manifold in Hilbert space through the mapping of local tensors \(A\) to physical states \(|\Psi(A)\rangle\), 
the tangent space of PEPS provides the foundation for variational ground state and excited states~\cite{Haegeman2013b,Vanderstraeten2015,Vanderstraeten2019a,Vanderstraeten2019b}. A tangent vector basis in the zero momentum sector can be written as
\begin{eqnarray}\label{eq:exst}
&&|\Phi(B;A)\rangle = B\cdot|\partial_A\Psi(A)\rangle=\sum_i |B_i\rangle\nonumber\\
 &=& \sum_i\sum_{abca'b'}^{s_{\alpha_i}s_{\beta_i}} B_{aba'b'}^{s_{\alpha_i}s_{\beta_i}} \frac{\partial}{\partial (A_{abc}^{s_{\alpha_i}} A_{a'b'c}^{s_{\beta_i}})} |\Psi(A)\rangle.
\end{eqnarray}
The gauge redundancy inherent in tensor-network representations makes the tangent space basis overcomplete. Impurity tensors of the form \( B = AX - XA \) produce null states within the tangent space and should be projected out for a well-defined basis of excited states~\cite{Vanderstraeten2019a,Vanderstraeten2019b}.

For a given Hamiltonian \(H\), the ground state energy \(E_0 = \langle \Psi(A)|H|\Psi(A)\rangle/\langle \Psi(A)|\Psi(A)\rangle\) is optimized using gradient-based methods, where the gradient \(g = \langle\partial_A \Psi(A)|H|\Psi(A)\rangle\)~\cite{Corboz2016,Vanderstraeten2016,Liao2019}. Consequently,  the energy of excited state is given by
\begin{equation}\label{eq:Ex}
    E_{\text{ex}} = \frac{\langle \Phi(B;A) | (H - E_0) | \Phi(B;A) \rangle}{\langle \Phi(B;A) | \Phi(B;A) \rangle}.
\end{equation}
The numerator in the excitation energy expression, Eq.~(\ref{eq:Ex}), must remain positive according to the variational principle, a fundamental requirement that must be strictly upheld during simulations. While it is possible to artificially enforce positivity by reducing the dimension of the tangent space when the numerator becomes negative, the approach is neither controllable nor reliable for accurately evaluating excitation gaps. In our simulations of chiral PEPS, we observe that the numerator in Eq.~(\ref{eq:Ex}) remains consistently positive and finite after a careful optimization of the ground state PEPS. 

Consequently, the primary challenge in managing the ``long-range'' correlation tails in chiral PEPS shifts to ensuring the stability of the denominator, \(\langle \Phi(B;A) | \Phi(B;A) \rangle = \sum_{ij} \langle B_i | B_j \rangle\), which corresponds to the structure factor of the zero-momentum correlation function of \(|B_i\rangle\). 
The key condition for stability lies in the decay behavior of the correlation functions. When correlations decay faster than \(r^{-2}\) with distance, the structure factor remains finite, avoiding divergences. Notably, previous studies have demonstrated that in chiral PEPS, the structure factor remains regular throughout the Brillouin zone~\cite{Chen2018,Hasik2022}. This behavior aligns with that of a gapped system rather than a critical one, providing evidence that the bulk gap of a chiral spin liquid (CSL) may be captured. However, a direct calculation of the energy gap offers a more definitive validation, which we now present.

\emph{$\mathbb{Z}_2$ chiral state. -- }
The \(\mathbb{Z}_2\) Kitaev model with chiral terms is governed by the Hamiltonian  
\begin{equation}\label{eq:z2model}  
    H = -J \sum_{\langle ij \rangle \alpha} \sigma_i^\alpha \sigma_j^\alpha - \kappa \sum_{ijk} \sigma_i^x \sigma_j^y \sigma_k^z,  
\end{equation}  
where \(\sigma_{x,y,z}\) are Pauli matrices. This Hamiltonian commutes with the flux operator \(W_p = \sigma_1^x \sigma_2^y \sigma_3^z \sigma_4^x \sigma_5^y \sigma_6^z\), which characterizes the \(\mathbb{Z}_2^{[1]}\) 1-form symmetry. The three-spin interaction term \(-\kappa \sum_{ijk} \sigma_i^x \sigma_j^y \sigma_k^z\) explicitly breaks time-reversal symmetry, yet the model remains exactly solvable. With small $\kappa$ (e.g., $\kappa/J = 0.2$ studied in this work), this model hosts a chiral ground state with 
non-Abelian anyonic excitations~\cite{Kitaev2006a}. The low-energy spectrum of this model was exactly solved explicitly in Ref.~\cite{Lahtinen2008}, providing a useful benchmark. 

We start from  a \(\mathbb{Z}_2^{[1]}\) 1-form symmetric initial PEPS  of virtual bond dimension \(D = 4\) and optimize the ground state energy of the model in Eq.~(\ref{eq:z2model}) at \(\kappa/J = 0.2\). Using the corner transfer matrix renormalization group (CTMRG) method~\cite{Nishino1996,Orus2009,Corboz2014} with an environment truncation dimension \(\chi = 80\), we achieve a variational ground state energy of \(E_0 = -1.6979J\), remarkably close to the exact result \(E_0 = -1.6980J\).   

\begin{figure}[b]
    \centering
    \includegraphics[width=\columnwidth]{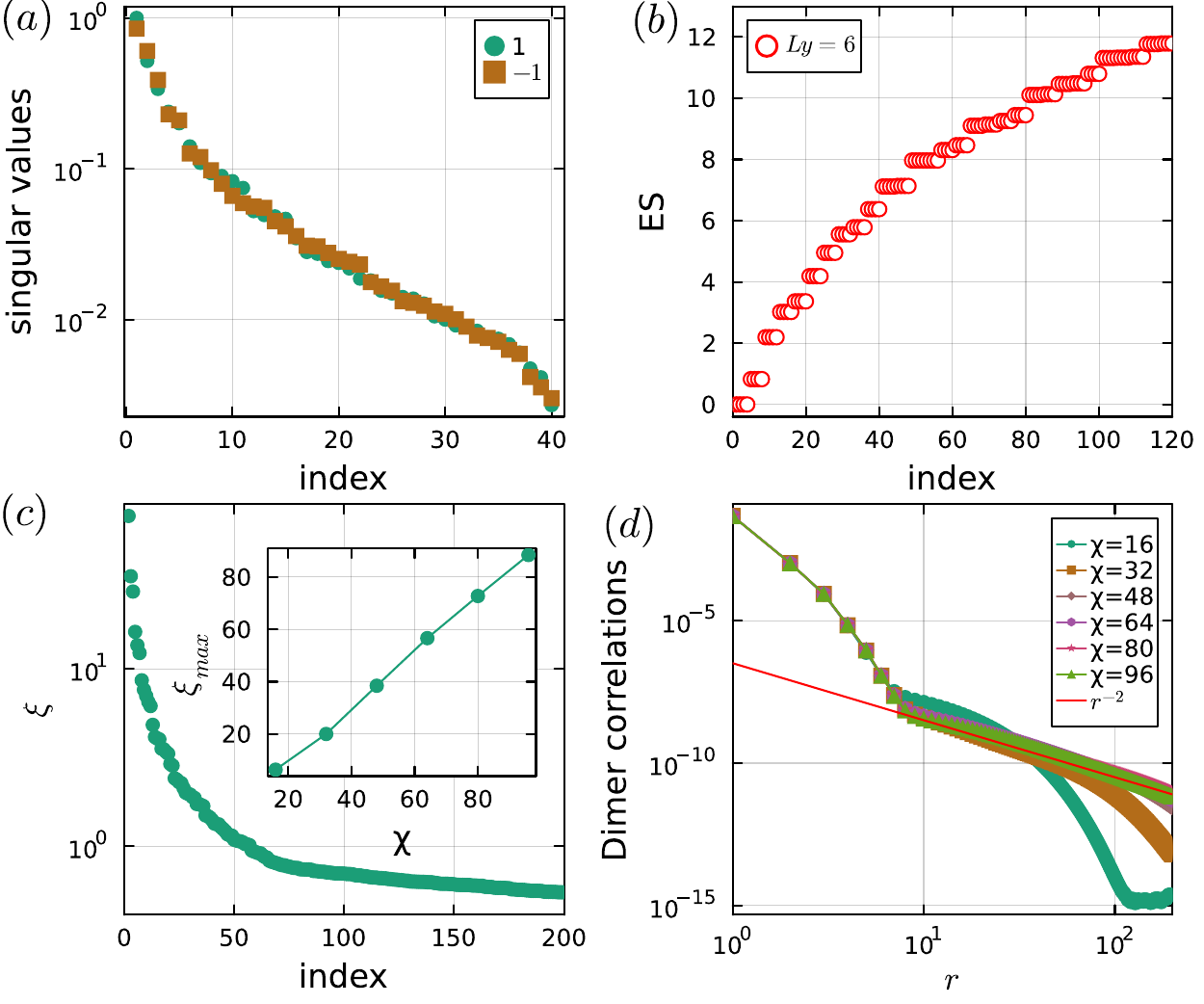}
    \caption{Properties of the ground state PEPS of the $\mathbb{Z}_2$ Kitaev model with chiral terms. (a) Singular values of  the corner matrix in different $\mathbb{Z}_2$ charge sectors.  (b) Entanglement spectrum. (c) Correlation lengths. The inset is $\chi$-dependent maximum correlation length. (d) Connected dimer-dimer correlation function $|\langle K(0)K(r)\rangle_c|$. 
    }
    \label{fig:Z2peps}
\end{figure}
Figure~\ref{fig:Z2peps} highlights the essential features of the ground-state PEPS for the $\mathbb{Z}_2$ model (\ref{eq:z2model}). During CTMRG process, we notice notoriously slow convergence. This is evident in Fig.~\ref{fig:Z2peps}(a), where the singular values of the corner matrix exhibit a smooth distribution. 
In Fig.~\ref{fig:Z2peps}(b), the entanglement spectrum is shown for an infinite cylinder with circumference \(L_y = 6\). The spectrum reveals characteristic degeneracies originating from local conservation laws of the flux~\cite{Yao2010,Shinjo2015}, imposed by the \(\mathbb{Z}_2^{[1]}\) 1-form symmetry. 

Figure~\ref{fig:Z2peps}(c) provides insights into the correlation lengths, extracted from the eigenvalue spectrum of the transfer matrix built with CTMRG environment tensors at bond dimension \(\chi = 80\). 
The inset of Fig.~\ref{fig:Z2peps}(c) shows how the maximum correlation length \(\xi\) diverges linearly with increasing \(\chi\).  
Finally, Fig.~\ref{fig:Z2peps}(d) illustrates the behavior of the connected dimer-dimer correlation function as a function of distance. At short distances, the correlation function decays exponentially, characterized by a correlation length \(\xi = 0.26\). At larger distances, however, the decay transitions into a power-law form, \(r^{-\alpha}\), with \(\alpha \simeq 2\) as \(\chi\) increases. 

To the best of our knowledge, \( r^{-2} \) represents the slowest decay behavior observed for the long-range correlation tails in chiral PEPS. Although a power-law decay \( r^{-2} \)  is marginal, its overall contribution is negligible. 
Specifically, the tail has an extremely small weight, less than \( w = 10^{-5} \), indicating that its effects would only become appreciable for system sizes on the order of  \( L \sim \exp(10^5) \). 
Consequently, we conclude that the long-range tail does not significantly impact the excitation spectrum in our simulations and does not introduce spurious gapless modes.

\begin{figure}[b]
    \centering
    \includegraphics[width=\columnwidth]{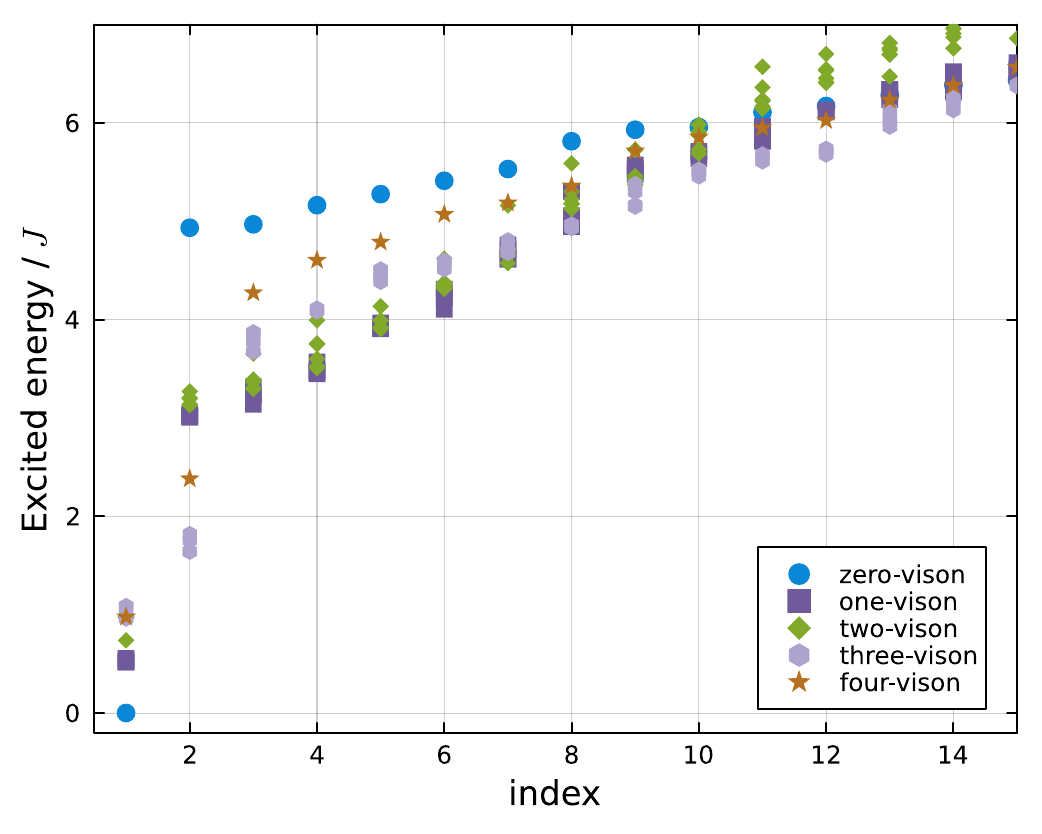}
    \caption{On-site energy levels of the $\mathbb{Z}_2$ Kitaev model with chiral terms using $\kappa/J=0.2$, where the location of excitation is fixed, taking the form Eq.~\eqref{eq:single_site}.
    }
    \label{fig:Z2energy_levels}
\end{figure}

Having established the ground-state properties, we now turn our attention to the calculation of excited states. Utilizing a variational approach within the local mode approximation, we study the excitations described by the states in Eq.~(\ref{eq:exst}). These excited states are constructed from the single-site excitation ansatz
\begin{equation}\label{eq:single_site}
|B_i\rangle =\includegraphics[width=0.45\columnwidth,valign=c]{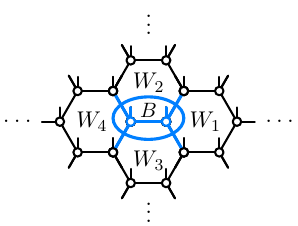}.    
\end{equation}
While the ground state inherently preserves the \(\mathbb{Z}_2^{[1]}\) 1-form symmetry, the single-site excited states in Eq.~(\ref{eq:single_site}) are also 1-form symmetric and distinguished by their local vison configurations~\cite{Tan2024,Tan2024a}. Specifically, the surrounding plaquette operators \(\hat{W}_{1,2,3,4}\) define the vison configurations, characterized by the expectation values \(W_{1,2,3,4} = \langle B_i|\hat{W}_{1,2,3,4}|B_i\rangle / \langle B_i|B_i\rangle\)~\cite{Tan2024,Tan2024a,Wang2024}. This classification reflects the underlying local conservation laws enforced by the 1-form symmetry, providing a structured framework for investigating excitations in the system.

These excited states fall into five distinct sectors: zero-vison, single-vison, two-vison, three-vison, and four-vison~\cite{Tan2024}. The zero-vison and four-vison sectors each exhibit a unique vison configuration. In contrast, the single-vison and three-vison sectors display four equivalent vison configurations, reflecting the underlying lattice symmetry. The two-vison sector features six possible configurations, five of which are equivalent, with one distinct configuration where \(W_1 = W_4 = -1\).  

In the nonzero-vison sectors, the 1-form symmetry enforces orthogonality for states between different sites~\cite{Baskaran2007, Tan2024}, ensuring that \(\langle B_i|H|B_j\rangle = \langle B_i|B_j\rangle = 0\) for \(i \neq j\). This orthogonality guarantees that not all correlation functions exhibit long-range tails in chiral PEPS. 
Consequently,  excited states are localized in non-zero vison sectors, and their energy is determined by the on-site value:  
\[
E = \frac{\langle B_0|H|B_0\rangle}{\langle B_0|B_0\rangle},
\]
where \(i = 0\) denotes $0$-th site. The on-site energy can be obtained by diagonalizing the corresponding eigen-problems. Figure~\ref{fig:Z2energy_levels} depicts the on-site energy levels across all vison sectors, including the zero-vison sector. Each sector features a spectral continuum, indicative of fractionalized excitations. Within this continuum, bound states emerge in the nonzero-vison sectors. 

For the zero-vison sector, excited states exhibit nonzero hopping amplitudes, \(\langle B_i|B_j\rangle \neq 0\) for \(i \neq j\), allowing delocalization and lowering the on-site energy.   
Using gradient-based optimization for the excited energy in Eq.~(\ref{eq:Ex}) , we  obtain the excitation gap of \(\Delta_{\text{zero-vison}} = 4.7J,\) in the zero-vison sector at zero momentum, approximately 18\% larger than the exact value of \(4J\) at the same momentum. Despite this slight overestimation, the gap of the \(\mathbb{Z}_2\) chiral spin liquid is determined by the single-vison sector, with a gap \(\Delta = 0.52J\)~\footnote{Analytically, it was shown that the smallest gap in the zero vison sector is \(2J\) at nonzero momenta~\cite{Lahtinen2008}.}, closely matching the exact result reported in Ref.~\cite{Lahtinen2008}.

\begin{figure}[b]
    \centering
    \includegraphics[width=\columnwidth]{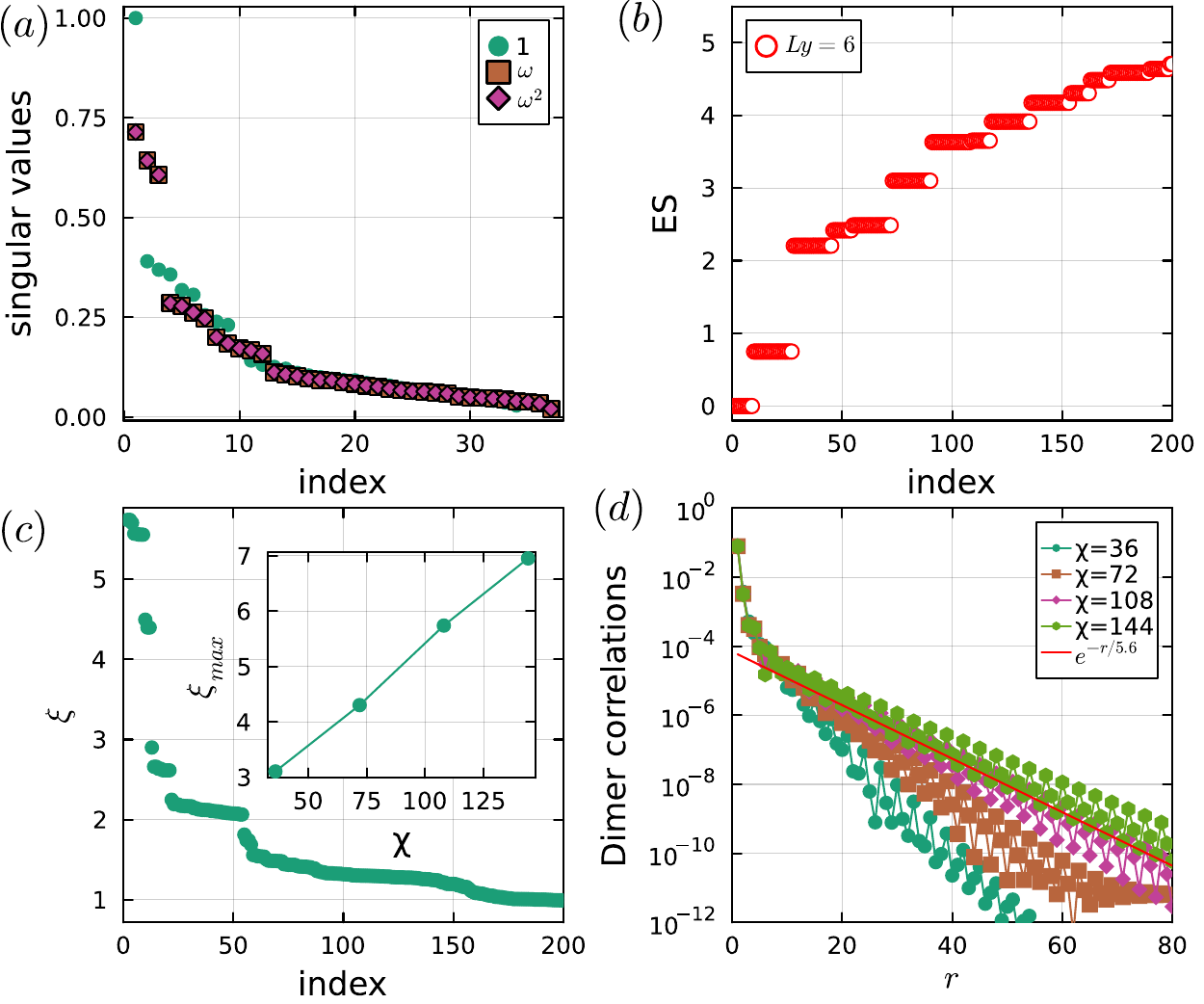}
    \caption{Properties of the ground-state PEPS for the \(\mathbb{Z}_3\) Kitaev model. (a) Singular values of the corner matrix in three $\mathbb{Z}_3$ charge sectors. (b) Entanglement spectrum. (c) Correlation lengths. (d) Connected dimer-dimer correlation function.
    }
    \label{fig:Z3peps}
\end{figure}    
\emph{$\mathbb{Z}_3$ chiral state. --}
The Kitaev model can be extended to the \(\mathbb{Z}_3\) case with the following Hamiltonian~\cite{Barkeshli2015}
\begin{equation}\label{eq:z3model}
    H = -J \sum_{\langle i,j\rangle \in \alpha} T_i^\alpha T_j^\alpha + {\rm H.C.},    
\end{equation}
with 
\begin{equation}
    T_x=\begin{pmatrix}
        0 &0& 1\\
        1&0&0\\
        0&1&0
    \end{pmatrix},
    T_y=\begin{pmatrix}
        0 &\omega^2& 0\\
        0&0&\omega\\
        1&0&0
    \end{pmatrix},
    T_z=\begin{pmatrix}
        1 &0& 0\\
        0&\omega&0\\
        0&0&\omega^2
    \end{pmatrix},
\end{equation}
with $\omega=e^{i2\pi/3}$.
The loop operator for each hexagon $p$ is defined as $W_p=(T_x^1T_y^2T_z^3T_x^4T_y^5T_z^6)^\dag$, which characterizes the $\mathbb{Z}_3^{[1]} $ 1-form symmetry in this model. This model lacks an exact solution. Prior studies, primarily relying on matrix product state (MPS) calculations on finite-size systems, have confirmed its chiral ground state~\cite{Chen2024,Eck2024}. However, the nature of its energy gap in the two-dimensional thermodynamic limit remains an unresolved question.

We optimize the \(\mathbb{Z}_3^{[1]}\) 1-form symmetric ground-state PEPS with bond dimension \(D = 6\) for the \(\mathbb{Z}_3\) model, using an environment bond dimension of \(\chi = 108\). The optimized ground-state energy per unit cell is \(E_0 = -2.9600J\). For comparison, exact diagonalization (ED) yields ground-state energies of \(E_0 = -3.013J\) for an 18-site torus and \(E_0 = -3.016J\) for a 24-site torus. This close agreement confirms the validity of the variational PEPS ground state.

Figure~\ref{fig:Z3peps} highlights the properties of the ground-state PEPS for the \(\mathbb{Z}_3\) Kitaev model, paralleling the results for the \(\mathbb{Z}_2\) case in Fig.~\ref{fig:Z2peps}. The singular value distribution in Fig.~\ref{fig:Z3peps}(a) is notably less critical than its \(\mathbb{Z}_2\) counterpart in Fig.~\ref{fig:Z2peps}(a). The entanglement spectrum in Fig.~\ref{fig:Z3peps}(b) exhibits a structure  consistent with previous DMRG results~\cite{Chen2024}, confirming the chiral topological nature.
Fig.~\ref{fig:Z3peps}(c) displays the correlation length for \(\chi = 108\), along with the (almost) linear scaling behavior of the maximal correlation length as a function of \(\chi\) (inset). Through explicitly computing dimer correlations, we have verified that the maximal correlation length corresponds to the dimer correlation. As shown in Fig.~\ref{fig:Z3peps}(d), the connected dimer-dimer correlation function for the \(\mathbb{Z}_3\) model exhibits a quick exponential decay at short distance.
In contrast to the \(\mathbb{Z}_2\) case, finite-\(\chi\) effects are considerably larger, preventing both the attainment of large correlation lengths and an accurate determination of the long-distance tail’s functional form. However, because the slowest exponential decay carries only a very small weight (\(w \sim 10^{-3}\)), its contribution can be safely neglected.

\begin{figure}[t]
    \centering
    \includegraphics[width=\columnwidth]{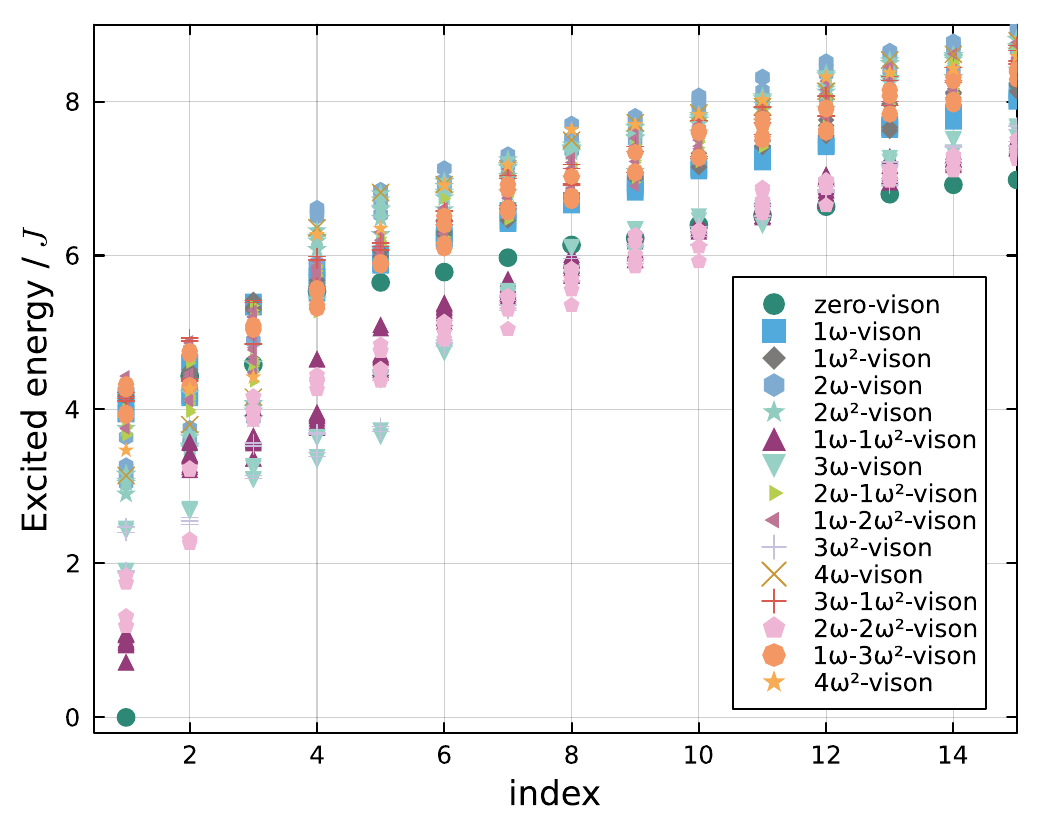}
    \caption{On-site energy levels of $\mathbb{Z}_3$ Kitaev model. Here each symbol corresponds to one specific vison number which may have multiple configurations.
    }
    \label{fig:Z3energy_levels}
\end{figure}   
We classify the single-site excited states \(|B_i\rangle\) based on the vison configurations of the surrounding plaquette operators, with eigenvalues \(W = 1, \omega, \omega^2\), where \(\omega = e^{\pm i2\pi/3}\). For the \(\mathbb{Z}_3\) model, the single-site states are categorized according to the eigenvalues of the four plaquette operators \(\hat{W}_{1,2,3,4}\). With \(D = 6\) and \(d = 3^2\), there are \(D^4d\) states, which can be classified into \(3^4 = 81\) distinct vison configurations. To simplify the analysis, we group these configurations into 15 distinct classes, as labeled in Fig.~\ref{fig:Z3energy_levels}, where the energy levels corresponding to each vison sector are also presented. 
Considering the hopping between different sites, the variationally optimized gap in the zero-vison sector at zero momentum is calculated to be \(\Delta_{\rm zero-vison} = 1.49J\). Consequently, the lowest system gap is determined to be \(\Delta = 0.71J\) in the \(1\omega\)-\(1\omega^2\)-vison sector, corresponding to the configuration \(W_1 = \omega\), \(W_{2,3} = 1\), and \(W_4 = \omega^2\).

For comparison, we have carried out ED computations on torus geometry with system size up to 24 sites. Using the $W_p$ operators as penalty, we found that the vison gap is \(0.38J \) (\(0.45J\)) for 18 (24)-site torus, although it is not easy to infer the vison content in the corresponding excited state. For the zero vison sector, it turns out that for a sufficiently large size, there are quasi-degenerate states in the ground state manifold below a much larger gap. For instance, on the 18-site torus, the level splitting in the ground state manifold is \(0.2J\) while the gap is \(4.5J\). While there could be strong finite size effect in the actual magnitude of gaps, the ED results do corroborate with the PEPS variational results in that the overall lowest excitation is indeed the vison excitation.

\emph{Conclusions. --}
This work establishes a robust framework for simulating gapped bulk excitations in chiral spin liquids using projected entangled-pair states. Leveraging a variational principle within the PEPS tangent space, we show that long-range correlation tails in chiral PEPS, which decay faster than \(r^{-2}\), do not generate spurious gapless bulk modes. Explicit calculations of the energy gaps for the \(\mathbb{Z}_2\) and \(\mathbb{Z}_3\) Kitaev models validate this approach. 

This study conclusively demonstrates the capability of PEPS to simulate gapped bulk excitations in chiral spin liquids, effectively addressing longstanding concerns regarding their applicability to chiral topological phases. Notably, as dictated by the no-go theorem for chiral PEPS, a gapped local Hamiltonian cannot serve as the exact parent Hamiltonian of the PEPS. Instead, we suggest that these parent Hamiltonians likely deviate slightly from the original Kitaev models by incorporating long-range interactions with negligible amplitudes.
Rethinking the correspondence between correlation length and energy gap, our work suggests that when dealing with chiral phases and their phase transitions, the finite correlation length scaling in PEPS~\cite{Corboz2018,Rader2018} may need refinement to distinguish ``gossamer'' tail from truly physical correlation. We leave further investigation in this direction to future works.

\emph{Acknowledgment --}
We thank Jize Zhao, Yongshi Wu, Hong-Hao Tu, Donghoon Kim, and Guang-Ming Zhang for useful discussions. We use the TensorKit package\footnote{https://github.com/Jutho/TensorKit.jl} to implement the block structure of the \(\mathbb{Z}_N\) gauge symmetry. This work is supported by  the National Natural Science Foundation of China (Grant No.~12474143), the National Key Research and Development Program of China (Grant No.~2021YFA1400400), and Shenzhen Fundamental Research Program (Grant No.~JCYJ20220818100405013  and JCYJ20230807093204010). J.-Y. C. was supported by  National Natural Science Foundation of China (Grant No.~12304186,~12447107), Innovation Program for Quantum Science and Technology 2021ZD0302100, Guangzhou Basic and Applied Basic Research Foundation (Grant No.~2024A04J4264), and Guangdong Basic and Applied Basic Research Foundation (Grant No.~2024A1515013065).
Part of the calculations reported were performed on resources provided by the Guangdong Provincial Key Laboratory of Magnetoelectric Physics and Devices, No.~2022B1212010008, and at CALMIP (grant 2024-P0677).
This work was also supported by the TNTOP ANR-18-CE30-0026-01 grant awarded by the French Research Council.

\bibliography{draft}

\end{document}